\documentclass[a4,12pt,onecolumn,oneside,notitlepage,final]{article}

\date{}
\usepackage{graphicx}
\usepackage[paperwidth=595pt,paperheight=842pt,
  top=70pt,right=70pt,bottom=70pt,left=70pt]{geometry}
\usepackage{caption}
\usepackage{titlesec}
\usepackage{microtype}
\titleformat*{\section}{\fontsize{14pt}{0pt}\selectfont\bfseries}
\titleformat*{\subsection}{\normalsize\bfseries}



\usepackage{amsmath, amssymb}
\begin{document}
\begin{flushleft}
{\fontsize{16pt}{0pt}\selectfont\textbf{Geometrical Shock Dynamics in a Noble-Abel stiffened gas medium}} \\
\vspace{12pt}
{\fontsize{12pt}{0pt}\selectfont { \textbf{M.I. Radulescu}}} \\
{\fontsize{12pt}{0pt}\selectfont { Mechanical Engineering, University of Ottawa, 161 Louis Pasteur, Ottawa, K1N 6N5, Canada}} \\
\vspace{12pt}
{\fontsize{12pt}{0pt}\selectfont {Matei Radulescu: matei@uottawa.ca}}
\end{flushleft}
\noindent
\textbf{Abstract} 
We derive the shock strength area rule for a Noble-Abel stiffened gas (NASG) equation of a state required in Whitham's geometrical shock dynamics approach to determine shock wave dynamics in dense gases, liquids and solids.  An exact formulation requiring the solution of an ordinary differential equation is provided.  Closed form solutions of various levels of approximations are also obtained as an expansion in shock wave strength.  The leading order approximation recovers the geometrical acoustic limit, while higher order approximations account for the medium's compressibility.  The exact shock strength area relation and the various order approximations are illustrated for shocks in liquid water. The simple closed form of the first order solution predicts the shock strength area rule up to shock pressures of approximately twice the stiffening pressure in water, i.e., approximately 1-2 GPa. 

\section{Introduction}
\label{subsec:1}
The present study extends Whitham's geometrical shock dynamics (GSD) method \cite{whitham} to treat the dynamics of shock waves in dense gases, liquids and solids.  The GSD method evolves the shock wave in a collection of ray tubes perpendicular to the shock surface and allows to predict shock surface discontinuities called \textit{shock-shocks}, the trace of transverse shocks.  It permits to calculate the shock dynamics in complex geometries at much smaller computational price than solving the entire flow-field.  It also provides analytical results in some canonical cases of shock diffraction, reflections and stability \cite{whitham}.  

Originally, Whitham's model was developed for shocks in ideal gases.  A few studies extended the method to water shocks \cite{best1991, cates1997,Wang_Eliasson_2012} using the Tait equation of state, to shocks in solids using the Mie-Gruneisen equation of state \cite{anand2022} and to magnetohydrodynamics (MHD) \cite{mostert_pullin_samtaney_wheatley_2017}.  The present study develops the method for a Noble-Abel Stiffened Gas \cite{tammann1912, lemetayer2016, radulescu2020compressible} . The Noble-Abel stiffened gas model offers a simple framework to study compressible flows and shock dynamics in dense gases, liquids and solids.  It relates the internal energy $e$ of the medium to the medium's pressure $p$ and specific volume $v$ :
\begin{equation} \label{eq:eNASG}
e(p,v)=\frac{p+\gamma p_\infty}{\gamma-1}\left(v-b \right)+q
\end{equation}
where $p_\infty$, $b$ and $q$ are fitting parameters and $\gamma$ is the ratio of specific heats.  This simple equation of state, first suggested by Tammann \cite{tammann1912} is a hybrid of the stiffened gas, or Tait equation of state ($b=0$), usually used to model compressible flows in liquids, particularly water, and the Nobel -Abel equation of state ($p_\infty=0$), usually used to treat dense gases.  It cures the thermodynamic inconsistency of the Tait equation of state, which does not permit to model temperature and the medium compressibility simultaneously \cite{radulescu2019}.   The explicit formulation of the internal energy on pressure and density makes this equation of state suitable for hydrodynamic simulations and its simplicity permits to use it in analytical approaches \cite{radulescu2020compressible}.  The correct thermodynamic behavior also permits to treat problems with rapid phase changes, such as cavitating flows. 

The GSD theoretical model requires two separate components. A physically derived area-shock strength rule is required for a given equation of state.   The method for advancement of the shock surface is then a simple kinematic problem independent of the physics; it applies equally to detonation waves and shock waves.   In the present study, we address the first component only, as the second is generic to any equation of state; different strategies for evolving the shock surface can be found in the literature \cite{henshaw_smyth_schwendeman_1986, schwendeman1993, cates1997, Schwendeman19991215, NOUMIR2015206}.  We wish to establish this area-shock strength rule for a NASG fluid. 

\section{The shock strength area rule in a NASG fluid}
\subsection{Rankine-Hugoniot shock jump relations}
The derivation of the shock strength - area rule requires first parametrization of the post shock state in terms of a single variable characterizing the shoick amplitude.  The shock jump relations for a NASG fluid were derived by Radulescu \cite{radulescu2020compressible} in terms of the shock Mach number:
\begin{equation}
M_s=\frac{D-u_1}{c_1}
\end{equation}
yielding:
\begin{equation}
\frac{p_2-p_1}{p_1+p_\infty}=\frac{2 \gamma \left(M_s^2-1  \right)}{\gamma+1}
\end{equation}
\begin{equation}
\frac{u_2-u_1}{c_1}\left( 1-\frac{b}{v_1} \right)^{-1}=\frac{2 M_s^2-1}{\left( \gamma+1 \right) M_s}
\end{equation}
\begin{equation}
\frac{ v_2 -b}{v_1 - b}=\frac{\left( \gamma -1 \right) M_s^2 +2}{\left( \gamma+1 \right) M_s^2}
\end{equation}
where the right hand sides are the same as for ideal gas.  They can thus be manipulated in a similar manner.  The density is obtained from the inverse of the specific volume, i.e.,
\begin{equation}
\rho=\frac{1}{v}
\end{equation}
and the sound speed for a NASG medium \cite{radulescu2020compressible} is given by
\begin{equation}
c^2=\gamma \frac{p+p_\infty }{\rho (1-\rho b}
\end{equation}

\subsection{Shock jump relations parametrized by the over-pressure $z$}
The shock jump equations are conveniently re-written in terms of a modified over-pressure $z$ defined as
\begin{equation}
z =\frac{p_2-p_1}{p_1+p_\infty} \label{eq:pz}
\end{equation}
and we can write
\begin{equation}
M_s=\left( 1 +\frac{\gamma+1}{2 \gamma} z \right)^\frac{1}{2} \label{eq:Mz}
\end{equation}
\begin{equation}
\frac{u_2-u_1}{c_1}\left( 1-\frac{b}{v_1} \right)^{-1}=\frac{z}{\gamma \left( 1 +\frac{\gamma+1}{2 \gamma} z \right)^\frac{1}{2}} \label{eq:uz}
\end{equation}
\begin{equation}
\frac{ v_2 -b}{v_1 - b}=\frac{1 +\frac{\gamma-1}{2 \gamma} z}{1 +\frac{\gamma+1}{2 \gamma} z} \label{eq:vz}
\end{equation}

\subsection{Shocks in water}
\begin{figure}
	\centering
	\includegraphics[width=0.5\textwidth]{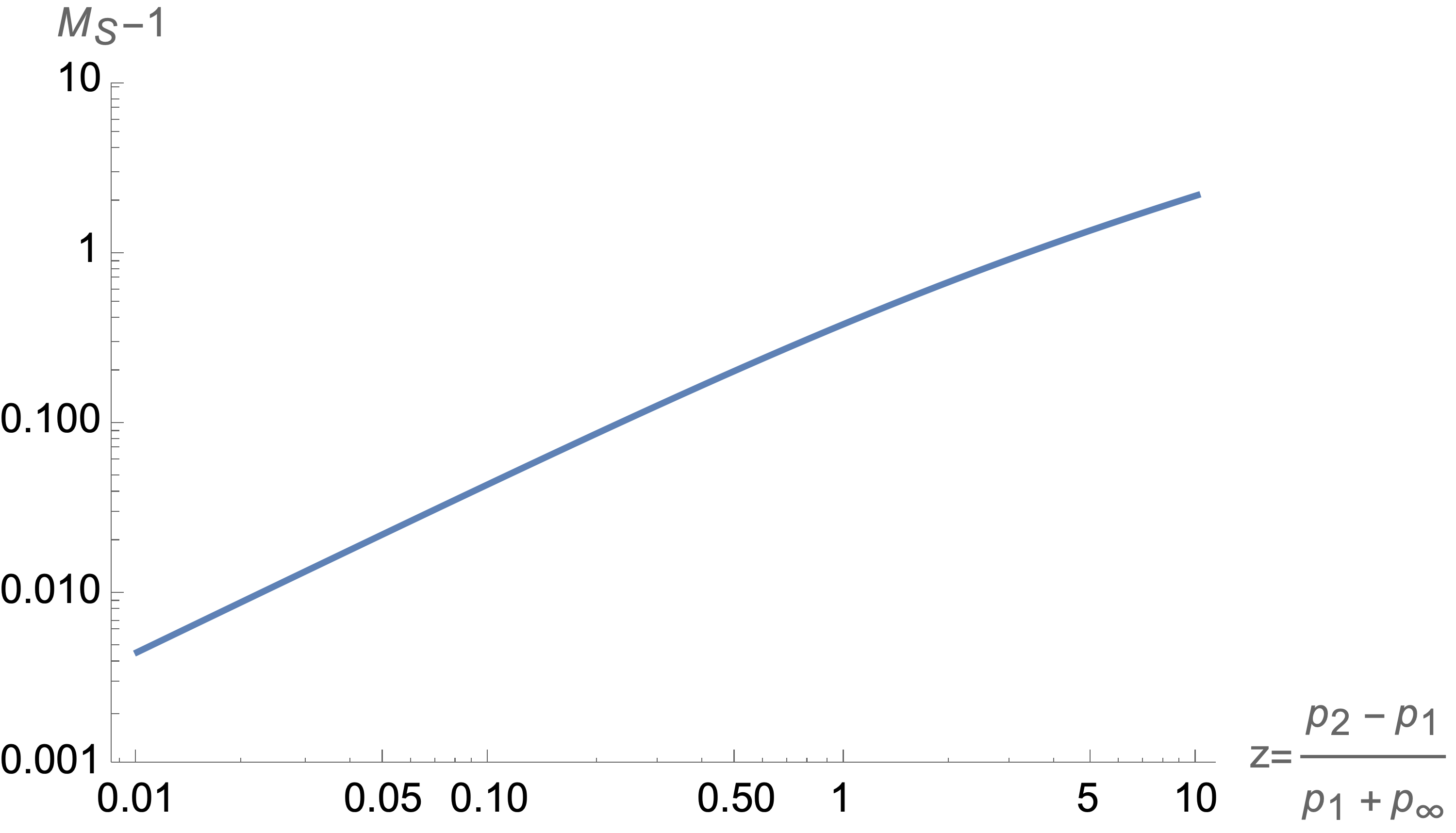}
	\caption{Mach number dependence on shock strength in water.}
	\label{fig:nasgm}
\end{figure}

\begin{figure}
	\centering
	\includegraphics[width=1.0\textwidth]{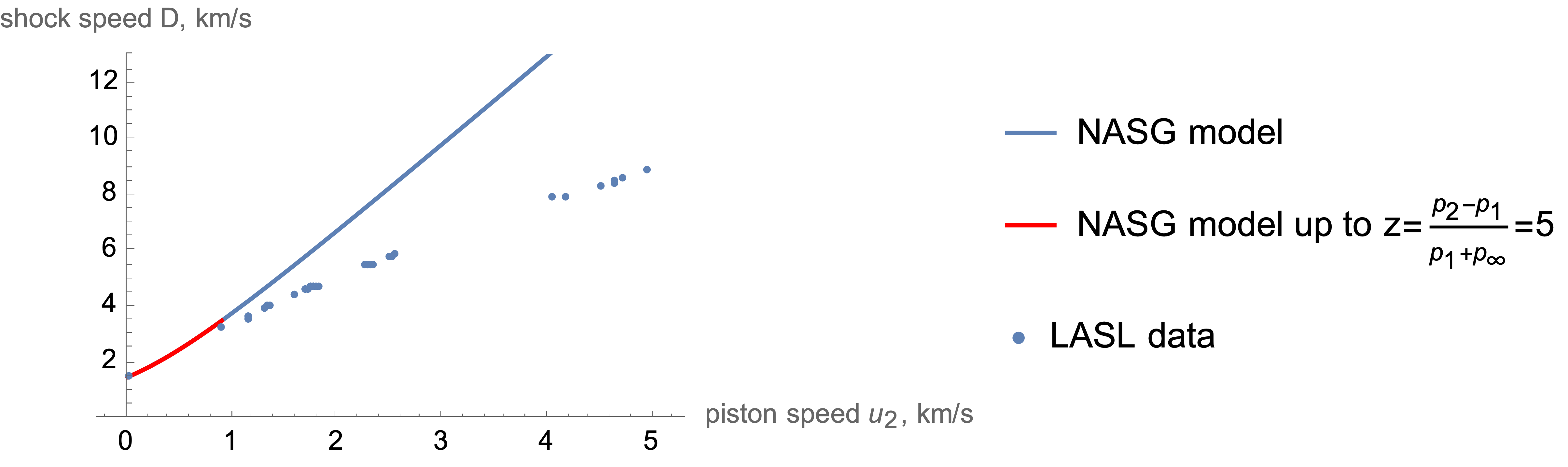}
	\caption{Shock speed dependence on piston speed in water and range of validity of the NASG equation of state.}
	\label{fig:nasgd}
\end{figure}

The NASG parameters for water have been fitted by LeMetayer \& Saurel \cite{lemetayer2016}.  They are listed in table \ref{tab_eos}.  The resulting dependence of Mach number on shock strength in shown in Fig.\ \ref{fig:nasgm} up to an overpressure of 5.  This is the limit of validity of the NASG model to capture shocks in water, as shown empirically from comparison with the strong shock data compiled in the LASL database \cite{lasl}, as shown in Fig.\ \ref{fig:nasgd}.
\begin{table} 
\begin{center}
\caption{NASG eos parameters for liquid water \cite{lemetayer2016}.}
\begin{tabular}{ccc}
$\gamma$ & $p_\infty$ (Pa) & $b$ (m$^3$/kg) \\
\hline
1.19& $7\times 10^8$ & $6.6 \times10^{-4}$ \\
\end{tabular}
\label{tab_eos}
\end{center}
\end{table}

\subsection{Whitham's characteristic rule}
The dependence of shock strength on ray tube area can be obtained by applying Whitham's general model, known as the \textit{characteristic rule} \cite{whitham}. The changes in the state behind the shock are transferred along the trajectory of a $C^+$ characteristic trailing closely the shock.  The justification is based on a small perturbation approach \cite{whitham}.  Changes in pressure, speed and local area along a $C^+$ characteristic is given by 
\begin{equation}
\mathrm{d}p+\rho c \mathrm{d}u =-\frac{\rho c^2 u}{u+c} \mathrm{d}(\ln A) \label{eq:cplus}
\end{equation}
When applying Whitham's characteristic rule, it is sufficient to recognize that each variable appearing in the $C^+$ equation \eqref{eq:cplus} is the post-shock state (subscript 2) and changes in pressure, for example can be expressed in terms of a shock parameter, Mach number or shock strength $z$, for example $\mathrm{d}p=(\mathrm{d}p / \mathrm{d}z) \mathrm{d}z$.  One obtains:
\begin{equation}
\frac{\mathrm{d}p}{\mathrm{d}z}+\rho c \frac{\mathrm{d}u}{\mathrm{d}z} =-\frac{\rho c^2 u}{u+c} \frac{\mathrm{d}\ln A}{\mathrm{d}z}
\end{equation}
This equation can be re-arranged as an explicit first order differential equation for $A(z)$, with the right hand side expressed explicitly in terms of $z$ alone. 
\begin{equation}
\frac{\mathrm{d}\ln A}{\mathrm{d} \ln z} = -z \left(  \frac{\mathrm{d}p}{\mathrm{d}z}+\rho c \frac{\mathrm{d}u}{\mathrm{d}z} \right) \frac{u+c}{\rho c^2 u} \label{eq:ODEfull}
\end{equation}
The derivatives appearing on the RHS of \eqref{eq:ODEfull} are evaluated from the shock jump relations for $p_2(z)$ \eqref{eq:pz} and for $u_2(z)$ \eqref{eq:uz} in a straightforward way.  The resulting ODE can be integrated numerically from an initial condition $A(z=z_0)=A_0$.  This provides the desired $A(z)$ rule.  The numerically obtained area-shock strength rule for shocks in water is shown in Fig.\ \ref{fig:zA}.  Note that the standard approach used by Whitham was to parametrize the area rule in terms of the shock Mach number.  Both treatments area equivalent but using the overpressure is preferred for weak shocks, where further simplifications can be obtained, as illustrated next.

\subsection{Weak shock approximations}
In condensed media, we expect shocks to be \textit{weak}, that is their Mach numbers to be close to unity and over-pressures satisfying $z\ll 1$.  A perturbation approach can thus be sought in this limit.  The jump relations can be expanded in powers of $z$, yielding:
\begin{equation}
M_s-1 =\frac{\gamma+1}{4 \gamma}z-\frac{(\gamma+1)^2}{32 \gamma^2}z^2+O(z^3)
\end{equation}
\begin{equation}
\frac{u_2-u_1}{c_1}\left( 1-\frac{b}{v_1} \right)^{-1}=\frac{z}{\gamma}-\frac{\gamma+1}{4 \gamma^2}z^2+O(z^3)
\end{equation}
\begin{equation}
\frac{ \rho_2 -\rho_1}{\rho_1}  \left( 1-\frac{b}{v_1} \right)^{-1} =\frac{z}{\gamma} - \left(\frac{\gamma-1}{2} + \frac{b}{v_1}  \right)  \left(\frac{z}{\gamma}\right) ^2+O(z^3)
\end{equation}
\begin{equation}
\frac{ c_2 -c_1}{c_1} =\left(\frac{\gamma-1}{2} + \frac{b}{v_1}  \right) \frac{z}{\gamma} + \frac{\gamma^2-1}{8\gamma^2} z^2+O(z^3)
\end{equation}

The right hand side of \eqref{eq:ODEfull} can thus be written as a power series in  $z$.  Retaining only the first two terms for presentation simplicity, one obtains:
\begin{equation}
\frac{\mathrm{d}\ln A}{\mathrm{d} \ln z} = -2 + \left( \frac{\gamma-3}{2 \gamma} + \frac{2}{\gamma}\frac{b}{v_1} \right) z + O(z^2)
\end{equation}
with boundary condition $A(z_0)=A_0$.  
The leading order solution yields:
\begin{equation}
\frac{A}{A_0} = \left(\frac{z}{z_0}\right)^{-2}
\end{equation}
This corresponds to the geometrical acoustics result discussed by Whitham \cite{whitham}, with the corrections provided by the new definition of the overpressure $z$ involving the stiffening pressure $p_\infty$.  

The next order correction is also readily obtained:  
\begin{equation}
\frac{A}{A_0} = \left(\frac{z}{z_0}\right)^{-2} \exp\left( \left( \frac{\gamma-3}{2 \gamma} + \frac{2}{\gamma}\frac{b}{v_1} \right) (z-z_0)\right)
\end{equation}
and one can easily proceed to higher order. 

\begin{figure*}
	\centering
	\includegraphics[width=1.0\textwidth]{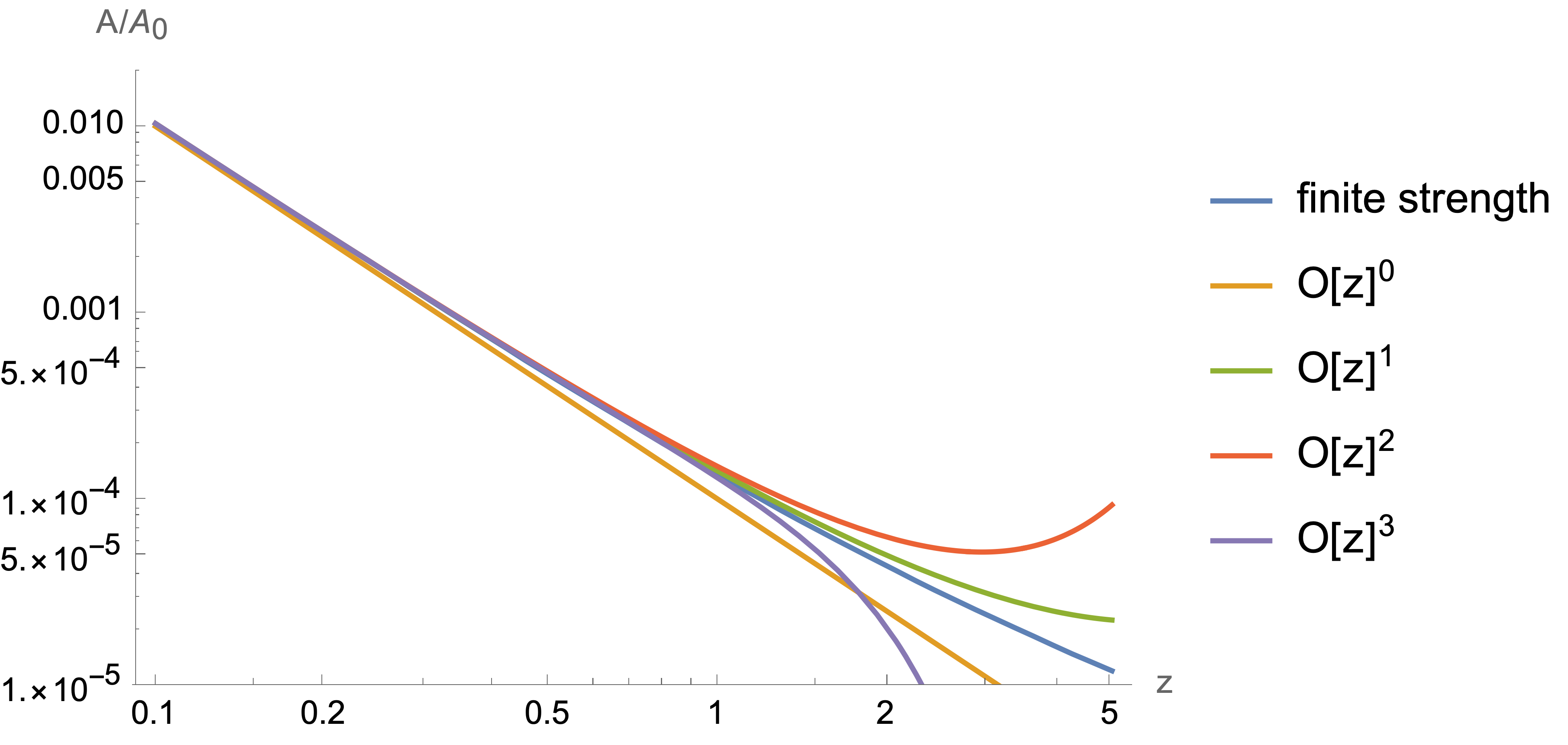}
	\caption{Shock strength - area dependence using the NASG parameters for water.}
	\label{fig:zA}
\end{figure*}

The relation between the ray tube area $A$ and the shock over-pressure is shown in Fig.\ \ref{fig:zA} for shocks in water.  The parameters of the NASG eos for water are taken from \cite{lemetayer2016} and reproduced in Table \ref{tab_eos}.  The solutions obtained for various order of approximation using the asymptotic theory for weak shocks is compared with the results obtained by integrating \eqref{eq:ODEfull} directly.   For comparison purposes, we have calculated these by taking the same reference point for sufficiently weak shocks by taking $z_0=0.0001$.  The range of comparison extends to a shock overpressure of 5, for which the shock Mach number is 2.36 and the post shock (piston speed) $u_2$ is approximately 1000 m/s.  This is approximately the range of validity of the NASG eos to treat shocks in water, as shown in Fig.\ \ref{fig:nasgd}, where we compare the shock jump conditions with the LASL data for shock Hugoniot in water. 

The higher order solutions improve on the geometrical acoustics solution at finite shock strengths.  The first order solution is notably closer to the full non-linear result outside the expected range of small $z$.  General agreement extends to approximately $z \simeq 1$ (shock Mach numbers of approximately 1.4). For shocks in water, this falls in the GPa range.
   
\section{Conclusion}
We have derived a simple shock strength- ray tube area rule for shocks in a NASG fluid following Whitham's characteristic rule.  The exact dependence can be obtained numerically by integrating a single ODE.  For weak shocks, simple closed form solutions were obtained.  These permit to seek analytical solutions in the same class of problems discussed by Whitham for ideal gases.  

\bibliographystyle{unsrt} 
\bibliography{references}
%
%
%
%

\end{document}